# Scattering of quasiparticle of spin-triplet pairs in diffusive superconductor-ferromagnetic nanowire-superconductor junction

Lin He

Department of Physics, Beijing Normal University, Beijing, 100875, People's Republic of China

Abstract:

 We analyze the proximity effect of superconductor/ferromagnet nanostructures and point out that the scattering of quasiparticle of spin-triplet pairs by local magnetic moments leads to large resistance peak slightly below the superconducting transition temperature of the superconductor. Both the temperature and the magnitude of the resistance peak decrease with increasing magnetic field, which agree quite well with the experimental results reported in [Nature Phys. **6**, 389 (2010)].

E-mail: helin@bnu.edu.cn

The Kondo effect and superconductivity are two of the most widely studied many-body phenomena in solid-state physics. When present simultaneously, these two effects are usually expected to be competing physical phenomena. In a conventional superconductor the pairs are spin-singlet pairs in which one spin points up and the other down. The local magnetic moments tend to align the spins of the electron pairs which often results in a strongly reduced transition temperature of the superconductor. Consequently, when a spin-singlet superconductor is placed in contact with a ferromagnet, it is expected that the pair correlations penetrate into the ferromagnet only a few nanometers $\xi_F \approx \sqrt{\frac{\hbar D}{k_B T_{Curie}}}$ [1,2], with Planck's constant divided by $2\pi$ $\hbar$, the electron diffusion constant $D$, Boltzmann's constant $k_B$, and Curie temperature of the ferromagnet $T_{Curie}$. Whereas, recent experiments reveal the observation of supercurrent through a high Curie temperature ferromagnet over several hundreds of nanometers [2-6], which is similar to the long-range proximity observed in superconductor/normal metal structures [7]. The unusually superconductivity in ferromagnets was demonstrated arising from spin-triplet pairing induced at the interface between a spin-singlet superconductor and a ferromagnet with inhomogeneous magnetization near the interface [2-6,8-10].

The spin-triplet proximity effect raises several theoretically interesting challenges. The interplay between superconductivity and ferromagnetism leads to unexpected phenomena observed in experiments. In a recent paper, Jian *et al.* reported a strikingly large and sharp resistance peak appears near the superconducting transition temperature of the electrodes $T_C^S$ in a study of single-crystal ferromagnetic cobalt nanowires sandwiched between superconducting electrodes [3]. This unexpected peak cannot be easily explained on the basis of the existing theoretical models, including a non-equilibrium charge imbalance mechanism [11-13] and spin-accumulation mechanism [14,15].

In this paper, we show that the scattering of quasiparticle of spin-triplet pairs by ferromagnet leads to large and sharp resistance peak slightly below $T_C^S$. According to our analysis, with increasing magnetic field, the magnitude of the resistance peak is suppressed as it moves to lower temperatures. These results quantitatively agree with that observed experimentally [3].

Fig. 1 shows a superconductor/ferromagnet(S/F) structure with a local inhomogeneity of the magnetization in the ferromagnetic wire near the S/F interface. Both the spin-singlet and spin-triplet components are induced in the ferromagnet wire due to the proximity effect [8]. The singlet component penetrates into the ferromagnet over a short length of $\xi_F$. The penetration length of the triplet component is equal to $\xi_\varepsilon \approx \sqrt{\frac{\hbar D}{\varepsilon}}$, where the energy $\varepsilon$ is of the order of temperature $k_B T$. Therefore, $\xi_\varepsilon$ is of the some order as that for the penetration of Cooper pairs into a normal metal $\xi_N \approx \sqrt{\frac{\hbar D}{k_B T}}$. The magnitude of the spin-triplet pairs depends on the inhomogeneity near the S/F interface. In the case of a homogeneous magnetization of the ferromagnetic wire the triplet pairing cannot be induced [8].

Jian *et al.* carried out several control experiments to explore the origin of the resistance peak slightly below $T_C^S$ [3]. Fig. 2a shows three typical S/F structures measured in the experiments [3]. For case *I*, the length of the ferromagnetic nanowire $L \leq 2\xi_\varepsilon$, there is no resistance peak below $T_C^S$ (Fig. 1b in Ref. [3]). For case II and III, $L > 2\xi_\varepsilon$, a sharp peak slightly below $T_C^S$ is observed in the temperature dependencies of resistance measurements (Fig. 2, Fig. 4, and Fig. 5a in Ref. [3]). Additionally, this unexpected peak was not seen in non-magnetic Au nanowires even when the length of the Au nanowire is larger than $2\xi_N$ [7]. Therefore, we can conclude that the ferromagnet plays an essential role in the appearance of resistance peak. Recent experiments reveal the observation of Kondo effect in a quantum dot configuration contacted with two ferromagnetic leads [16] and even in ferromagnetic atomic contacts [17]. We demonstrate here that the resistance peak is evidence for Kondo effect induced by quasiparticles of spin-triplet Cooper pairs through the ferromagnetic nanowire.

Though the spin-triplet Cooper pair, differs from an electron, is not point-like particle, we treat the pair as a whole, *i.e.* it is treated as a "particle" with 2*e* and *S* = 1.

Let us assume that the quasiparticles of spin-triplet Cooper pairs penetrated into the ferromagnetic nanowire and screened the local magnetic moments. This is similar as the screening of local magnetic moments by the conduction electrons [18,19]. Then the resistance of the nanowire can be expressed as [18,19]

$$R(T) \approx R_0(1 + 4J\rho_F \ln(\frac{k_B T}{E_F})) \quad . \qquad (1)$$

Here $R_0$ is the resistance of the nanowire slightly above $T_C^S$, $J$ is the antiferromagnetic exchange coupling between quasiparticles of spin-triplet Cooper pairs and local magnetic moments, $\rho_F$ is the quasiparticle density of spin-triplet pairs at the Fermi energy, $E_F$ is the Fermi energy of the nanowire. For the case $L >> \xi_F$, one can obtain $N_P(T) \approx N_S(T)/\sinh(\alpha_w)$ at the S/F interface, where $N_P(T)$ and $N_S(T)$ are the quasiparticle density of spin-triplet and spin-singlet pairs penetrated into the nanowire respectively, $\alpha_w$ is the angle characterizing the rotation of the magnetization near the S/F interface [8]. In Bardeen-Cooper-Schrieffer (BCS) superconductor [20,21],

$$N_S(T) \approx N_N \left| \text{Re} \frac{E}{\sqrt{E^2 - \Delta^2(T)}} \right| , \qquad (2)$$

where $N_N$ is the density of states above $T_C^S$, $\Delta(T)$ is the energy gap of BCS superconductor. In Eq. (2), all quasiparticle energies should be taken into account by integrating over all energies, and a Fermi-Dirac distribution should be used to take account of the temperature. According to Eq. (1) and Eq. (2), the resistance induced by the quasiparticles at energy $E$ can now be written as

$$R(T) \approx R_0(1 + 4JN_N/\sinh(\alpha_w) \left| \text{Re} \frac{E}{\sqrt{E^2 - \Delta^2(T)}} \right| \ln(\frac{k_B T}{E_F})). \qquad (3)$$

Near $T_C^S$, $\Delta(T) \approx 1.74 (1-T/T_C^S)^{1/2}$ [20]. With this approximation, we have

$$R(T) \approx R_0(1 + a/\sqrt{T - T_P}). \qquad (4)$$

Here $\alpha = 4JN_N(T_C^S)^{0.5} E \ln(\frac{k_B T_C^S}{E_F})/(1.74\sinh(\alpha_w))$, and $T_P = T_C^S(1 - \frac{E^2}{1.74^2})$.

Obviously, the resistance of the nanowire begins to increase at $T_C^S$ and reaches a maximum at $T_P$ with taking into account the scattering of quasiparticle of spin-triplet by local magnetic moments, as shown by the schematic curve in Fig. 2b. Eq. (4) produces infinite resistance at $T = T_P$. This incorrect result arises from the assumption of Eq. (2), which assumes the density of spin-singlet pairs is infinite at $E = \Delta(T)$.

Fig. 3 shows the resistance as a function of temperature slightly below $T_C^S$ for an individual 80 nm Co nanowire with $L = 1.5$ μm [3]. The expected spin-singlet penetration length $\xi_F$ in Co nanowire is estimated as 1.8-4.5 nm [3], i.e., $L \gg \xi_F$. Therefore, the experimental data should be described by Eq. (4) by integrating over all energies. Here we use Eq. (4) as a phenomenological equation to fit the experimental result. The solid curve in Fig. 3 is the computed result of Eq. (4) with $T_P = 4.44$ K. Obviously, Eq. (4) can give a quantitatively description of the experimental results. The peak $T_P = 4.44$ K obtained by Eq. (4) also quantitatively agrees with the experimental result, $T_P \sim 4.4$ K [3]. With considering $T_C^S \sim 5.2$ K, $E$ is estimated as 0.67 K in zero applied magnetic field.

According to Eq. (4), we obtain $T_P = T_C^S (1 - \frac{E^2}{1.74^2})$. It indicates that the temperature of the resistance peak is proportional to the superconducting transition temperature of the electrodes $T_C^S$, which decreases with increasing the applied magnetic fields (Fig. S1 in Ref. [3]). Fig. 4 shows the dependencies of $T_P$ and $T_C^S$ on the magnetic field, obtained from Fig. 2b of Ref. [3]. Obviously, both $T_P$ and $T_C^S$ show similar dependent trends on the magnetic field. The absolute difference ($T_P - T_C^S$) seems to be constant. It indicates that the value of $E$ is not constant and it depends on the magnetic field.

The increase of resistance arising from Kondo effect can be written as $\Delta R(T) = R(T) - R_0 \approx 4R_0 J N_P(T) \ln(\frac{k_B T}{E_F}))$ according to Eq. (1). Obviously, the value of $\Delta R(T)$ is proportional to the density of spin-triplet pairs $N_P(T)$ through the nanowire. With assuming that the local inhomogeneity of the magnezation in the vicinity of the S/F interface arises only from domain walls near the interface, then the inhomogeneity of the magnetization and the density of the spin-triplet pairs induced at the interface decrease with increasing the magnetic field [8]. Consequently, the magnitude of the resistance peak slightly below $T_C^S$ will decrease with increasing the magnetic field and the resistance peak will even disappear with sufficient large field. At present it is difficult to make a quantitative comparison of our theory to experiments for that there exists several possible mechanisms to create the inhomogeneity near the S/F interface, including domain walls at the interface, local magnetic impurities/defects, surface spin-glasses and so on. However, the magnitude of resistance peak decreasing with increasing magnetic field predicted in our model qualitatively agrees with the experimental results, as shown in Fig. 2 and Fig. 5a reported in Ref. [3].

Recently, the competition between triplet superconductivity and Kondo effect was studied by Koga and Matsumoto [22]. According to their analysis, the Kondo effect strongly depends on spin and orbital degrees of freedom of Cooper pairs. In this work, we just use the usual free electron perturbative Kondo result and take into account the quasiparticle density of states of superconductor. It is amazing that the calculated results quite agree with all the aspects of the recent experimental results [3]. Further work should be done to explain how scattering of triplet cooper pairs by magnetic impurities can lead to Kondo like resistivity behavior, as described by Eq. (1).

In conclusion, we have shown that the interaction between spin-triplet pairs and local magnetic moments leads to large resistance peak slightly below $T_C^S$. The temperature and the magnitude of the resistance peak decrease with increasing magnetic field, which agree quite well with the experimental results reported in Ref. [3].

This work was financially supported by National Natural Science Foundation of China (No. 11004010).

Figure caption:

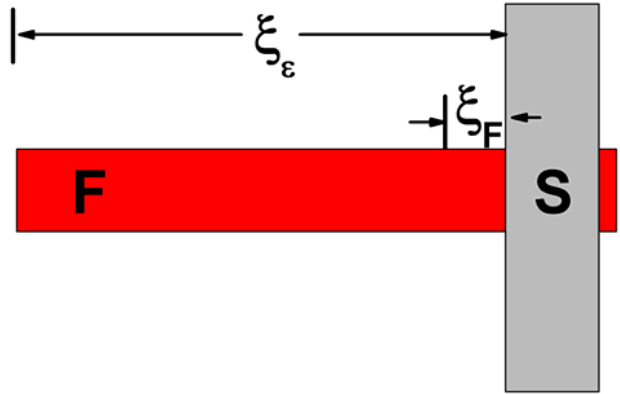

**FIG. 1**. Schematic view of the superconductor/ferromagnet (S/F) structure under consideration. $\xi_\varepsilon$ and $\xi_F$ are the penetration length of spin-triplet pairs and spin-singlet pairs into the ferromagnetic wire respectively.

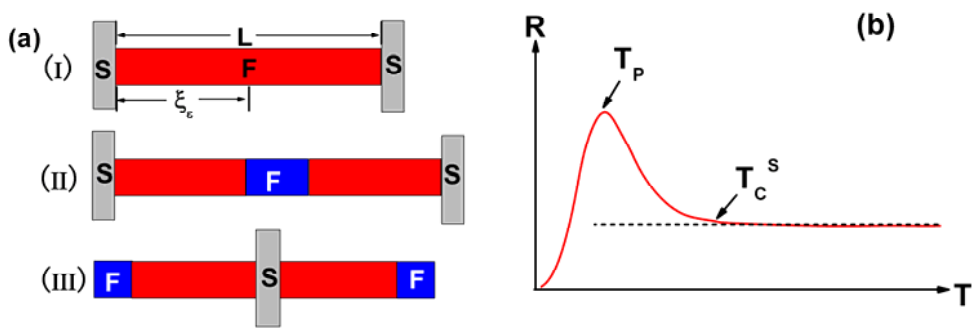

**FIG. 2**. (a): Schematic view of three typical S/F structures measured in Ref. [3]. (b): Typical temperature dependencies of resistance for case (II) and (III) S/F structures.

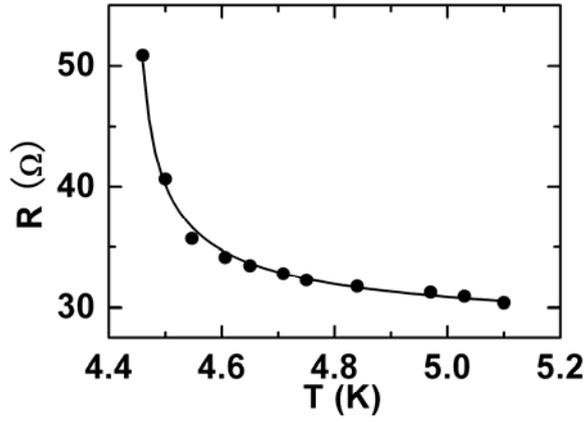

**FIG. 3**. The solid circles are the resistance as a function of temperature slightly below $T_C^S$ for an individual 80 nm Co nanowire with $L$ = 1.5 μm [3]. The solid curve is the computed result of Eq. (4) with $T_P$ = 4.44 K.

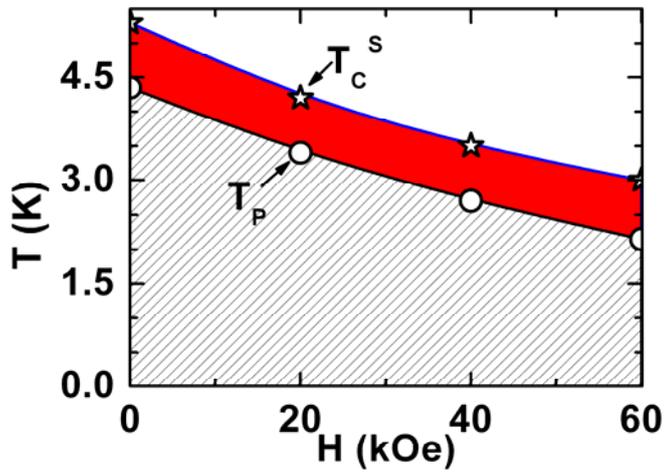

**FIG. 4**. The dependencies of $T_P$ and $T_C^S$ on the magnetic field, obtained from the Fig. 2b of Ref. [3].